\input{aipcheck}

\documentclass[
    ,final            % use final for the camera ready runs
  ]
  {aipproc}

\layoutstyle{8x11single}

\newcommand\lsim{\mathrel{\rlap{\lower4pt\hbox{\hskip1pt$\sim$}}
    \raise1pt\hbox{$<$}}}
\newcommand\gsim{\mathrel{\rlap{\lower4pt\hbox{\hskip1pt$\sim$}}
    \raise1pt\hbox{$>$}}}

\newcommand{\beq}{\begin{equation}}
\newcommand{\eeq}{\end{equation}}
\newcommand{\bea}{\begin{eqnarray}}
\newcommand{\eea}{\end{eqnarray}}
\newcommand{\bem}{\begin{pmatrix}}
\newcommand{\eem}{\end{pmatrix}}

\newcommand{\TeV}{\ \mathrm{TeV}}
\newcommand{\GeV}{\ \mathrm{GeV}}

\usepackage{color}
\definecolor{gray}{rgb}{0.5,0.5,0.5}

\begin{document}

\title{Challenging the minimal supersymmetric SU(5) model\footnote{Talk given by Borut Bajc.}}

\classification{12.10.-g, 12.60.Jv
%<Replace this text with PACS numbers; choose from this list:
%                \texttt{http://www.aip..org/pacs/index.html}>
}

\author{Borut Bajc}{
  address={J.\ Stefan Institute, 1000 Ljubljana, Slovenia}
  ,altaddress={Department of Physics, University of Ljubljana, 1000 Ljubljana,
Slovenia} 
}

\author{St\'ephane Lavignac}{
  address={Institut de Physique Th\'eorique\footnote{Laboratoire
de la Direction des Sciences de la Mati\`ere du Commissariat \`a l'Energie 
Atomique et Unit\'e de Recherche associ\'ee au CNRS (URA 2306).} , CEA-Saclay, 
F-91191 Gif-sur-Yvette Cedex, France}
}

\author{Timon Mede}{
  address={J.\ Stefan Institute, 1000 Ljubljana, Slovenia}
}

\begin{abstract}
We review the main constraints on the parameter space of the minimal renormalizable
supersymmetric SU(5) grand unified theory. They consist of the Higgs mass, proton decay, 
electroweak symmetry breaking and fermion masses. Superpartner masses are
constrained both from below and from above, giving hope for confirming 
or definitely ruling out the theory in the future. This contribution is based on Ref.~\cite{mi}.
\end{abstract}

\maketitle

%%%%%%%%%%%%%%%%%%%%%%%%%%%%%%%%%%%%%%%%%%%%
%% MAINMATTER
%%%%%%%%%%%%%%%%%%%%%%%%%%%%%%%%%%%%%%%%%%%%

\section{Introduction}

It is often claimed that the minimal supersymmetric SU(5) grand 
unified theory is ruled out. This statement is based on the analysis of Ref.~\cite{Murayama:2001ur}:
gauge coupling  unification constrains the colour Higgs triplet to be lighter than $\approx 3.6 \times 10^{15} \GeV$, while the
non-observation of proton decay requires it to be significantly heavier. If two 
massive colour triplets were available, like in the model with an extra vectorlike fundamental
representation~\cite{Babu:2012pb}, both constraints would be compatible, but in the minimal
supersymmetric SU(5) model the same triplet cannot be (relatively) light and heavy at the same time. 

This analysis relies on two implicit assumptions: 
\begin{itemize}
\item
the predictions of the model are not affected by the presence of non-renormalizable operators; 
\item
the soft supersymmetry breaking masses do not exceed the TeV scale, apart from the first two 
generations of sfermions, whose masses can be as large as $10 \TeV$.
\end{itemize}
How much does the conclusion of the above analysis depend on these assumptions? 
Allowing non-renormalizable operators can turn the theory back to life \cite{Bajc:2002pg}: on one side 
the masses of the colour octet and weak triplet inside the adjoint Higgs field can now differ,
which relaxes the unificaton constraint on the colour triplet mass~\cite{Bachas:1995yt,Chkareuli:1998wi};
on the other side the new terms allow for a much more flexible flavour structure in both the
sfermion~\cite{Bajc:2002bv} and the fermion sectors~\cite{Ellis:1979fg,EmmanuelCosta:2003pu},
thus relaxing the naive bound on the triplet mass from proton decay~\cite{Sakai:1981gr,Weinberg:1981wj,Hisano:1992jj}.
Of course this is possible at the expense of a large number of 
unknown parameters, which makes the theory less predictable. 

There is however an argument against the presence of such non-renormalizable terms. We know that the 
Planck-suppressed operator
\beq
\frac{\kappa}{M_{\rm Planck}}\,10_F^110_F^110_F^2\bar 5_F^l\, ,
\eeq
where $1, 1, 2, l$ are generation indices, must be further suppressed by $\kappa \lsim 10^{-7}$ in order 
to satisfy the strong proton decay bounds. This may tell us that all Planck-suppressed operators
enjoy some extra suppression. After all, we do not really know how gravity contributes to
the effective field theory below $M_{\rm Planck}$. It may well 
be that these operators are exponentially suppressed or even vanish exactly. This we will assume 
throughout this talk.

We are thus left with the second option, namely considering larger soft superpartner masses,
which by the way is also favoured by flavour physics constraints and by the mass of the recently
discovered Higgs boson. This is what this talk is about: we will require the superpartner masses
to fit all the experimental constraints in the minimal renormalizable supersymmetric SU(5) model.
All we assume are SU(5)-invariant boundary conditions for the soft terms, i.e. supersymmetry
must be broken above $M_{\rm GUT}$, as in supergravity. We will see that this is not a very
constraining assumption: in fact flavour-blind supersymmetry breaking mechanisms
like gauge mediation will not be available, since satisfying the various experimental
constraints requires a large splitting between sfermion generations. As we will show,
the third generation 
(as well as the higgsinos and the heavy Higgs bosons) must be very heavy ($\sim 10^{2-3} \TeV$)
in order not to destabilize the electroweak vacuum, while  the first two generations of sfermions
(and the gauginos) need to be lighter ($\sim 10 \TeV$) in order to be able to 
correct the SU(5) fermion mass relations. 

%%%%%%%%%%%%%%%%%%%%%%%%%%%%%%%%%%%%
\section{The minimal renormalizable supersymmetric SU(5) model}
%%%%%%%%%%%%%%%%%%%%%%%%%%%%%%%%%%%%

Let us start with a short description of the minimal renormalizable supersymmetric
SU(5) model. The Higgs sector consists of the adjoint $24_H$, whose vacuum expectation value 
\beq
\langle24_H\rangle=V\; {\rm Diag}\, (2,2,2,-3,-3)
\eeq
is responsible for the spontaneous gauge symmetry breaking 
SU(5) $\to$ SU(3)$_{\rm C} \times$SU(2)$_{\rm L} \times$U(1)$_{\rm Y}$.
This vev is found as a minimum of the scalar potential defined by the superpotential:
\beq
W_{24}\, =\, \frac{\mu}{2}\, \mbox{Tr}\, (24_H^2) + \frac{\lambda}{3}\, \mbox{Tr}\, (24_H^3)\ ,
\eeq
with
\beq
V=\frac{\mu}{\lambda}\ .
\eeq
Expanding the above superpotential around this minimum, the masses of the weak triplet
and colour octet components of the adjoint Higgs superfield are found to be:
\beq
\label{m3m8}
m_3=m_8= 5 \lambda V\ .
\eeq

The MSSM Higgs doublet pair lives in the $5_H \oplus \bar 5_H$ representation,
whose superpotential couplings are:
\beq
W_5=\bar 5_H(m+\eta 24_H)5_H\ .
\eeq
After breaking of the SU(5) gauge symmetry, the colour triplet and electroweak doublet components of the
Higgs fields $5_H \oplus \bar 5_H$ acquire the following masses:
\bea
M_T\, =\, m+2\eta V\ ,\\
m_D\, =\, m-3\eta V\ .
\eea
In order to recover the MSSM at low energy and to prevent too fast proton decay,
the doublet-triplet splitting is achieved by means of the fine-tuning:
\beq
m=3\eta V\, ,
\eeq
so that the mass of the heavy colour triplet is:
\beq
\label{mT}
M_T=5\eta V\, .
\eeq

Next, the Yukawa sector of the model is given by:
\beq
W_{\rm Yukawa}=Y^{10}_{ij}10_F^i10_F^j5_H+Y^5_{ij}\bar 5_F^i10_F^j\bar 5_H\ ,
\eeq
where $i,j$ are generation indices running from 1 to 3. This simple structure leads to the following 
predictions at the GUT scale:
\bea
M_U\, =\, M_U^T\, , \\
M_D\, =\, M_E^T\, .
\eea
The equality between the GUT-scale down quark and charged lepton masses leads
to particularly bad predictions for the first two generations and has to be corrected.
We will do that with the only resource we still have, i.e. supersymmetric threshold
corrections (for a recent analysis see for example Ref.~\cite{Enkhbat:2009jt} 
and references therein). 

Finally, the kinetic sector gives the heavy gauge bosons a mass:
\beq
\label{mV}
M_V=5g_5V\, .
\eeq
Comparing Eq.~(\ref{mT}) with Eq.~(\ref{mV}) and assuming perturbativity
(i.e. $\eta \lsim 1$), we thus obtain
\beq
\label{mTmV}
M_T\lsim M_V\, ,
\eeq
since $g_5 \approx 0.7$.

%%%%%%%%%%%%%%%
\section{Our assumptions}
%%%%%%%%%%%%%%%

It is time now to describe in detail the assumptions and inputs that we will use in our analysis:

\begin{itemize}

\item
we consider SU(5)-invariant, but otherwise arbitrary soft supersymmetry breaking terms;

\item
we insist on our electroweak vacuum to be the global minimum of the scalar potential. 
We will in particular impose the (strictly speaking not necessary) condition that~\cite{Casas:1995pd}
\beq
m_{H_u}^2>0
\eeq
at all energies between $M_{\rm GUT}$ and the electroweak symmetry breaking scale 
\beq
M_{\rm EWSB}\equiv\sqrt{\tilde m_{t_1}(M_{\rm EWSB})\tilde m_{t_2}(M_{\rm EWSB})}\ ,
\label{M_EWSB}
\eeq
where the matching between the MSSM and the SM is done;

\item
the fermion masses receive corrections dominantly from 1-loop gluino exchange;

\item
the $d=5$ proton decay operator is dominated by 1-loop wino exchange;

\item
in order to minimize the dangerous supersymmetric contributions to flavour-changing
processes, we assume the masses of the first two generations of 
sfermions to be approximately equal:
\beq
\tilde m_1\approx \tilde m_2 \;\;(\equiv \tilde m_{1,2})\, .
\eeq
\end{itemize}

%%%%%%%%%%%%%%%%%%%%%%%%
\section{Fermion mass corrections}
%%%%%%%%%%%%%%%%%%%%%%%%

As already mentioned, in the minimal renormalizable supersymmetric SU(5) model,
the GUT-scale mass relations are only corrected by 1-loop finite supersymmetric
threshold corrections (for a recent work see for example \cite{Enkhbat:2009jt} and references therein).
We will only consider the corrections to down quark masses, which are enhanced
(compared with the corrections to charged lepton masses) by the strong coupling
constant:
\beq
\label{deltam}
\delta m_i=-\frac{2 \alpha_3}{3\pi}\frac{m_{\tilde g}v}{\tilde m_{1,2}^2}\left(A_i\cos{\beta}-\mu h_i\sin{\beta}\right)
I_1(m_{\tilde g}^2/\tilde m_{1,2}^2)\ ,
\label{mi_thresholds}
\eeq
where $m_i$ and $A_i$ are the down quark masses and the associated A-terms,
and the loop function $f$ is given by:
\beq
I_1(x)\equiv\frac{1-x+x\log{x}}{(1-x)^2}\ .
\eeq
Comparing the SU(5) predictions for fermion masses (obtained by running
the GUT-scale values down to low energy) with the experimental values
one finds that these corrections need to be positive for the down
quark and negative for the strange quark. 
This means that the term proportional to $\mu$ cannot dominate Eq.~(\ref{deltam}). 

So the dominant part comes from the $A_i$'s, which however are bounded
by vacuum stability constraints~\cite{Casas:1995pd}:
\beq
A_i=a_ih_i\sqrt{3} \left( 2\tilde m_{1,2}^2+m_{H_d}^2 \right)^{1/2} , \quad |a_i|\leq 1\ .
\label{Ai_constraint}
\eeq
Since the required corrections to the down and strange quark masses
are larger than the masses themselves, large $A_i$ terms are needed.
Then the only way to maintain vacuum stability while avoiding a suppression
of the corrections~(\ref{deltam}) by large squark or gluino masses is to assume
\beq
m_{H_d}\, \gg\, \tilde m_{1,2}\, ,\, m_{\tilde g}\ .
\label{vacuum_stability}
\eeq
Such a heavy $H_d$ gives a large 1-loop contribution to the hypercharge D-term
(by contrast it only receives small contributions from the sfermions, due to the SU(5)
boundary conditions on soft masses), which yields tachyons in the superpartner
spectrum unless it is compensated for by the contribution of $H_u$. We will
enforce this solution by assuming
\beq
  m^2_{H_u} (M_{\rm GUT})\, =\, m^2_{H_d} (M_{\rm GUT})\, \equiv\, m^2_{H}\, . 
\eeq
The large value of $m_{H_u}$ in turn tends to dominate the renormalization group
equations of the stop masses and drive them negative at low energy.
To avoid tachyons, the soft mass parameter $m_{10_3}$ (and to a lesser
extent $m_{\bar 5_3}$) must be large, hence the third generation sfermions
must be heavier than the first
two generation sfermions (and gauginos, according to Eq.~(\ref{vacuum_stability})).

For a fixed value of $m_{H_d}$, the values of the A-terms are bounded
by Eq.~(\ref{Ai_constraint}), which together with the requirement that
Eq.~(\ref{mi_thresholds}) accounts for the measured down quark masses
implies an upper limit on the soft masses $\tilde m_{1,2} \sim m_{\tilde g}$.

%%%%%%%%%%%%%
\section{Proton decay}
%%%%%%%%%%%%%

Here the tendency for the soft masses is opposite than in the previous section:
the higher the  supersymmetry breaking scale, the longer the proton lifetime.
Our estimate of the proton decay lifetime in the minimal renormalizable
supersymmetric SU(5) model is (using the hadronic matrix elements of
Ref.~\cite{Aoki13}):
\beq
  \tau (p \rightarrow K^+ \bar \nu)\ \approx\ 2 \times 10^{32}\, \mbox{yrs}\,
    \left( \frac{\tilde m_{1,2}}{10 \TeV} \right)^2
    \left( \frac{1/3\, I_1 (1/9)}{m_{\tilde w} / \tilde m_{1,2}\, I_1 (m^2_{\tilde w} / \tilde m^2_{1,2})} \right)^2
    \left( \frac{2 \tan \beta}{1 + \tan^2 \beta} \right)^2 \left( \frac{M_T}{10^{17} \GeV} \right)^2 .
\label{eq:tau_p_estimate}
\eeq
This is to be compared with the experimental constraint
$\tau (p \rightarrow K^+ \bar \nu) > 2.3 \times 10^{33}$~yrs ($90 \%$~C.L.)~\cite{p_K+nubar}.
Heavy colour triplet and first two generation sfermions are favoured,
as well as a small value of $\tan \beta$. 
With $M_T = 10^{17} \TeV$ and $\tan \beta = 1.7$, the experimental bound
is saturated by Eq.~(\ref{eq:tau_p_estimate}) for $\tilde m_{1,2} \approx 30 \TeV$.

%%%%%%%%%%%%%%%%%%%%%%%%%%
\section{Gauge coupling unification constraints}
%%%%%%%%%%%%%%%%%%%%%%%%%%

The colour triplet and superpartner masses are actually not independent
of each other when gauge coupling unification constraints are taken into account. 
One can solve numerically the 2-loop renormalization group equations (RGEs)
for gauge couplings with the top quark Yukawa coupling evolved at 1-loop only 
(the other Yukawa couplings are neglected, including the bottom quark one,
since the proton decay constraint forces $\tan \beta$ to be small). 
The matching between the SM and the MSSM is performed at the scale $M_{\rm EWSB}$,
while the one between the MSSM and the SU(5) theory is done at $M_{\rm GUT}$.
The mass splittings in the MSSM and in the GUT spectra are taken into account
with 1-loop threshold corrections. Putting everything together, one arrives
at the following constraints:
\bea
\label{mTsol}
\frac{M_T}{M_{\rm GUT}}&=&\exp{\left[\frac{5\pi}{6}\left(-\alpha_1^{-1}+3\alpha_2^{-1}-2\alpha_3^{-1}\right)_{2-loop}(M_{\rm GUT})\right]}
\left(\frac{m_3}{m_8}\right)^{5/2}\nonumber\\
&\times&\left(\frac{m_{\tilde w}}{m_{\tilde g}}\right)^{5/3}
\prod_{i=1}^3\left(\frac{m_{\tilde Q_i}^4}{m_{\tilde u^c_i}^3m_{\tilde e^c_i}}\frac{m_{\tilde L_i}^2}{m_{\tilde d^c_i}^2}\right)^{1/12}
\left(\frac{m_{\tilde h}^4m_A}{M_{\rm EWSB}^5}\right)^{1/6} ,  \\
\label{mVsol}
\frac{\left[M_V^2(m_3m_8)^{1/2}\right]^{1/3}}{M_{\rm GUT}}&=&
\exp{\left[\frac{\pi}{18}\left(5\alpha_1^{-1}-3\alpha_2^{-1}-2\alpha_3^{-1}\right)_{2-loop}(M_{\rm GUT})\right]}\nonumber\\
&\times&\left(\frac{M_{\rm EWSB}^2}{m_{\tilde w}m_{\tilde g}}\right)^{1/9}
\prod_{i=1}^3\left(\frac{m_{\tilde u^c_i}m_{\tilde e^c_i}}{m_{\tilde Q_i}^2}\right)^{1/36} ,
\eea
where we kept the dependence on $m_3$ and $m_8$ separately, although
$m_3 = m_8$ holds in the minimal renormalizable SU(5) model.
The $(\alpha_i)_{2-loop}(M_{\rm GUT})$ are calculated using only SM/MSSM
RGEs without thresholds, so they do not unify. Notice that the matching
scales $M_{\rm EWSB}$ and $M_{\rm GUT}$ drop out at leading order as they
must \footnote{In particular, the combination $M_{\rm GUT}\, \exp{\left[\frac{5\pi}{6}\left(-\alpha_1^{-1}+3\alpha_2^{-1}-2\alpha_3^{-1}\right)_{2-loop}(M_{\rm GUT})\right]}$
does not depend on $M_{\rm GUT}$ at the 1-loop level.}.

In the approximation where all superpartners lie at a single scale $M_{\rm EWSB}$
and gauge couplings are evolved with 1-loop RGEs, the colour triplet and heavy
gauge boson masses scale as~\cite{Hisano:1992jj}:
\bea
& M_T\, \propto\, M_{\rm EWSB}^{5/6}\ ,  \\
& M_V\, \propto\, M_{\rm EWSB}^{-2/9}\ .
\eea
This is good to memorize. It tells us that the partial proton lifetime due
to $d=5$ operators goes essentially as
\beq
\tau_p(d=5)\propto M_T^{22/5} ,
\eeq
i.e. it increases even faster with the mass of the heavy mediator than the usual gauge exchange 
mode of non-supersymmetric theories.
On the other hand, the dependence of $M_V$ on $M_{\rm EWSB}$ is rather weak,
and according to Eq.~(\ref{mVsol}), the perturbativity constraint~(\ref{mTmV})
can be satisfied by decreasing the colour octet and weak triplet masses,
i.e. by decreasing the parameter $\lambda$ in Eq.~(\ref{m3m8}).

Reintroducing the mass splittings in the superpartner spectrum in Eqs.~(\ref{mTsol})
and~(\ref{mVsol}), one finds that the most important contribution to $M_T$ (resp. $M_V$)
comes from the higgsino mass (resp. from gaugino masses)~\cite{Hisano:1992jj}.

%%%%%%%%%%%%%%%%%
\section{The Higgs boson mass}
%%%%%%%%%%%%%%%%%

Since the third generation sfermions are constrained to be very heavy, the electroweak
symmetry breaking scale $M_{\rm EWSB}$ defined by Eq.~(\ref{M_EWSB}) is necessarily
much higher than the  electroweak scale $v = 174 \GeV$. 
In this case, the standard MSSM Higgs mass formula cannot be applied:
one must decouple the heavy superpartners at the scale $M_{\rm EWSB}$
and evolve the Higgs quartic coupling $\lambda$ with the Standard Model RGE
down to the electroweak scale, where the Higgs mass is given at leading order
by the relation $m_h = \sqrt{2 \lambda v^2}$.

Neglecting the splittings between superpartner masses, the Higgs mass determines a unique 
relation between $\tan{\beta}$ and $M_{\rm EWSB}$ through the boundary condition
$\lambda(M_{\rm EWSB}) = (g^2 + g^{\prime 2}) (M_{\rm EWSB}) \cos^2 2 \beta / 4$.
Solving the 1-loop Standard Model RGE for the Higgs quartic coupling, one obtains
\beq
\tan{\beta}=1.7\;\;\;\Rightarrow\;\;\;M_{\rm EWSB}= 411 \TeV\, .
\eeq
These are the values we will consider in the following.

%%%%%%%%%%%%%%%%%%%%%%%%%%%%%
\section{The GUT and heavy superpartner spectrum}
%%%%%%%%%%%%%%%%%%%%%%%%%%%%%

We already learnt that $m_{H_d}$ must be large in order for the needed threshold
corrections to down quark masses to be consistent with the absence of
dangerous charge and colour breaking minima. We have also seen that
imposing the equality of $m_{H_u}$ and $m_{H_d}$ at $M_{\rm GUT}$
prevents the generation of a large hypercharge D-term that would induce
tachyons in the spectrum. It follows that both $m_{H_u}$ and
$m_{H_d}$ are large over an important energy range (in fact, only
$m_{H_u}$ runs sizably between $M_{\rm GUT}$ and $M_{\rm EWSB}$),
which has an important impact on the running of the third generation
sfermion and Higgs soft masses. Since we consider small values of
$\tan \beta$, for which $\lambda_b \ll \lambda_t$, this impact is most
significant for the stop and $H_u$ soft masses:
\bea
m_{H_u}^2(m)&=&m_{H}^2\ -\ \frac{2\tilde m_{10_3}^2+m_{H}^2}{2}\,
\left[\, 1 - \left(\frac{m}{M_{\rm GUT}}\right)^{3\bar\lambda_t^2(m)/4\pi^2}\, \right]\, ,  \\
\tilde m_{u_3}^2(m)&=&\tilde m_{10_3}^2\ -\ \frac{2\tilde m_{10_3}^2+m_{H}^2}{3}\,
\left[\, 1 - \left(\frac{m}{M_{\rm GUT}}\right)^{3\bar\lambda_t^2(m)/4\pi^2}\, \right]\, ,  \\
\tilde m_{Q_3}^2(m)&=&\tilde m_{10_3}^2\ -\ \frac{2\tilde m_{10_3}^2+m_{H}^2}{6}\,
\left[\, 1 - \left(\frac{m}{M_{\rm GUT}}\right)^{3\bar\lambda_t^2(m)/4\pi^2}\, \right]\, ,
\eea
in which the subleading Yukawa and gauge contributions have been
neglected\footnote{Let us recall that $m_{\tilde g}, m_{\tilde w} \ll m_{H_d}$,
or in terms of GUT-scale parameters $M_{1/2} \ll m_{H}$.}, and
\beq
\bar\lambda_t^2(m)\equiv\frac{\int_{\log{m}}^{\log{M_{\rm GUT}}}\lambda_t^2(m')d\log{m'}}{\log{(M_{\rm GUT}/m)}}\ .
\eeq

\begin{figure}
\includegraphics[height=.3\textheight]{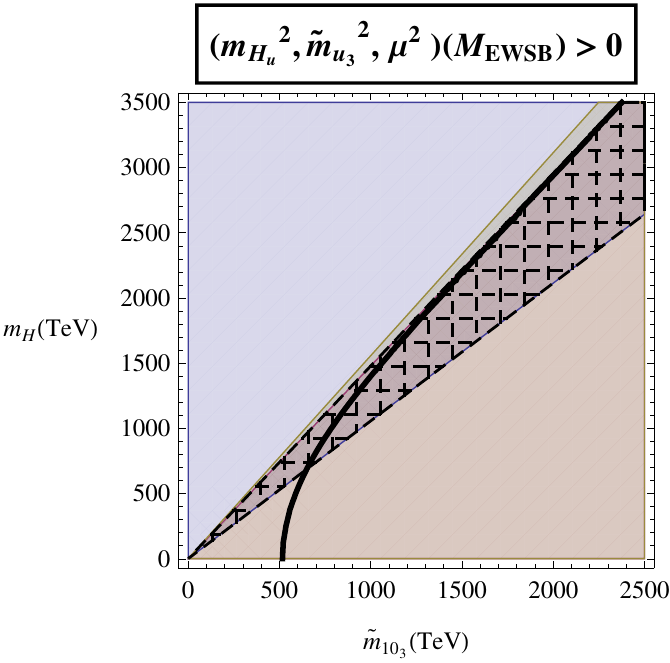}
\caption{\label{fig1} The allowed region in the ($\tilde m_{10_3}$, $m_{H}$)
plane corresponds to the intersection of the black curve with the hatched area
(here for $\tan{\beta}=1.7$, $M_{\rm EWSB} = 411 \TeV$ and $M_{\rm GUT} = 10^{17} \GeV$).
See text for details.}
\end{figure}

Now we can look for the region of the parameter space satisfying the constraints: 
\bea
\label{hu}
& m_{H_u}^2\ >\ 0\ ,\\
\label{Q3}
& 
\tilde m^2_{Q_3} ,\ 
\tilde m_{u_3}^2\ >\ 0\ , \\
\label{mu}
& |\mu|^2\, =\, \frac{m_{H_d}^2-m_{H_u}^2\tan^2{\beta}}{\tan^2{\beta}-1}\ \, >\ 0\ ,
\eea
at the electroweak symmetry breaking scale~(\ref{M_EWSB}), which in the absence
of a significant stop mixing is well approximated by:
\beq
\label{mewsb}
M_{\rm EWSB}\, \equiv\, \sqrt{\tilde m_{Q_3}(M_{\rm EWSB})\tilde m_{u_3}(M_{\rm EWSB})}\ .
\eeq
Condition~(\ref{mu}) is equivalent to saying that electroweak symmetry breaking is possible, 
while condition~(\ref{hu}) ensures the absence of charge and colour breaking minima
deeper than our electroweak vacuum.
Since, for a given value of $\tan \beta$, $M_{\rm EWSB}$ is determined by the Higgs
boson mass, Eq.~(\ref{mewsb}) implies 
a unique relation between $m_{H} \equiv m_{H_u}(M_{\rm GUT})=m_{H_d}(M_{\rm GUT})$
and $\tilde m_{10_3} \equiv \tilde m_{Q_3}(M_{\rm GUT}) = \tilde m_{u_3}(M_{\rm GUT})$,
which is represented by the black curve on Fig.~\ref{fig1}. The light blue region in the top left, the 
magenta and grey regions in the lower right satisfy the inequalities~(\ref{hu}), (\ref{Q3}) and~(\ref{mu}), respectively, 
while the hatched region is the portion of the parameter space allowed by all three constraints. 
Notice that relaxing the inequality~(\ref{hu}) would just allow the points along
the black curve below the shaded region, towards smaller values of $m_{H}$,
not enlarging significantly the parameter space. 

Thus, at the level of approximation described above, the region of the
parameter space of the minimal renormalizable supersymmetric SU(5) model
compatible with all experimental constraints depends on a single parameter,
say $\tilde m_{10_3}$. 
Taking for example:
\beq
\label{m103}
\tilde m_{10_3}=2000 \TeV\, ,
\eeq
we find:
\bea
&& m_{\tilde h}\, =\, \mu\, =\, 677 \TeV\, ,  \\
&& m_A\, =\, \sqrt{(\mu^2+m_{H_d}^2)(1+1/\tan^2{\beta})}\ =\, 3508 \TeV\, ,
\eea
where $m_A$ is the mass of the (approximately degenerate) heavy MSSM
Higgs bosons. Plugging these parameters into Eq.~(\ref{mTsol}) and neglecting
the mass splittings within the first two generations of sfermions, and using the
approximate relation $m_{\tilde w} / m_{\tilde g} \simeq (\alpha_2 / \alpha_3) (m_{\tilde g})$,
we obtain for $M_{\rm GUT}= 10^{17} \GeV$ and $\tilde m_{1,2} = m_{\tilde g} = 11 \TeV$:
\bea
& M_T\, =\, 4.0\times10^{17} \GeV\, ,  \\
& [M_V^2(m_3m_8)^{1/2}]^{1/3}\, =\, 8.5\times10^{15} \GeV\, .
\eea
The exact value of the sfermion and gaugino masses turn out to have a small 
effect on the colour triplet mass, while the combination of masses
$[M_V^2(m_3m_8)^{1/2}]^{1/3}$ is only weakly dependent on sfermion masses.

%%%%%%%%%%%%%%%%%%%%%%%%
\section{The ``light'' superpartner spectrum}
%%%%%%%%%%%%%%%%%%%%%%%%

\begin{figure}
\includegraphics[height=.3\textheight]{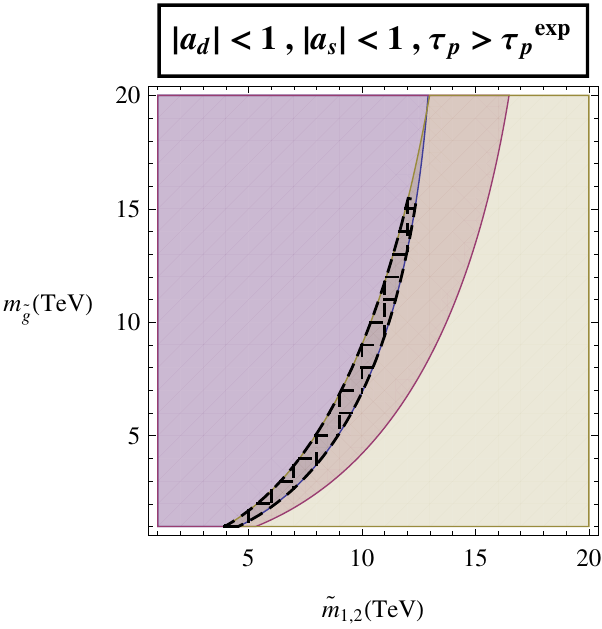}
\caption{\label{fig2}The allowed region in the ($\tilde m_{1,2}$, $m_{\tilde g}$) plane
corresponds to the hatched area (here for $\tan{\beta}=1.7$, $M_{\rm EWSB} = 411 \TeV$,
$M_{\rm GUT} = 10^{17} \GeV$ and $\tilde m_{10_3}=2000 \TeV$). See text for details.}
\end{figure}

We are now in a position to determine also the ranges of the gaugino and first two
generation sfermion masses for which the predicted proton lifetime and the
fermion masses are consistent with experiment. The four different regions in 
the ($\tilde m_{1,2}$, $m_{\tilde g}$) plane on Fig.~\ref{fig2}
represent from left to right the following constraints: the purple satisfies both 
$|a_d|<1$ and $|a_s|<1$, the hatched corresponds to the allowed region (all three 
constraints satisfied), the brown means both $|a_s|<1$ and $\tau_p>\tau_p^{exp}$, 
while the extreme right region satisfies only the proton decay bound.

%%%%%%%%%%%%
\section{Conclusions}
%%%%%%%%%%%%

We have identified a region in the parameter space of the minimal renormalizable
supersymmetric SU(5) model that is consistent with all experimental and
theoretical constraints:
gauge coupling unification, the measured charged fermion and Higgs boson masses,
the absence of charged and colour breaking vacua and the experimental limit
on the proton lifetime. The analysis has been simplified by making suitable approximations. 
Some single points in the allowed parameter space have been studied with a better
precision, and the results are compatible with the estimates presented in this talk.

Let us finish with a short comment about neutrino masses and dark matter. 
In the absence of additional multiplets like SU(5) singlets, the only possibility here seems
to include bilinear R-parity violating terms 
(for a review see for example Ref.~\cite{Barbier:2004ez}). After a generalized doublet-triplet 
splitting, there are enough parameters to fit the neutrino masses and mixings. In this 
context, the only dark matter candidate is a light, order GeV or less \cite{Takayama:2000uz} 
gravitino.

%%%%%%%%%%%%%%%%%%%%%%%%%%%%%%%%%%%%%%%%%%%%
%% Sample figure:
%%
%% The option [height=...] scales the picture to the given height,
%% without it it would be printed at its nominal size
%%%%%%%%%%%%%%%%%%%%%%%%%%%%%%%%%%%%%%%%%%%%

%\begin{figure}
%  \includegraphics[height=.3\textheight]{golfer}
%  \caption{Picture to fixed height}
%\end{figure}

%%%%%%%%%%%%%%%%%%%%%%%%%%%%%%%%%%%%%%%%%%%%
%% SAMPLE TABLE
%%
%% Shows the use of \tablehead and \tablenote
%% macros
%%%%%%%%%%%%%%%%%%%%%%%%%%%%%%%%%%%%%%%%%%%%

%\begin{table}
%\begin{tabular}{lrrrr}
%\hline
%  & \tablehead{1}{r}{b}{Single\\outlet}
%  & \tablehead{1}{r}{b}{Small\tablenote{2-9 retail outlets}\\multiple}
%  & \tablehead{1}{r}{b}{Large\\multiple}
%  & \tablehead{1}{r}{b}{Total}   \\
%\hline
%1982 & 98 & 129 & 620    & 847\\
%1987 & 138 & 176 & 1000  & 1314\\
%1991 & 173 & 248 & 1230  & 1651\\
%1998\tablenote{predicted} & 200 & 300 & 1500  & 2000\\
%\hline
%\end{tabular}
%\caption{Average turnover per shop: by type
%  of retail organisation}
%\label{tab:a}
%\end{table}

%%%%%%%%%%%%%%%%%%%%%%%%%%%%%%%%%%%%%%%%%%%%%%%%
%% BACKMATTER
%%%%%%%%%%%%%%%%%%%%%%%%%%%%%%%%%%%%%%%%%%%%%%%%

\begin{theacknowledgments}
We would like to thank Kaladi Babu, Ilia Gogoladze, Miha Nemev\v sek, Goran Senjanovi\'c and 
Zurab Tavartkiladze for discussion and correspondence. This work has been supported in part
by the Slovenian Research Agency and by the French Agence Nationale de la Recherche
under Grant ANR 2010 BLANC 0413 01.
BB and SL thank the Galileo Galilei Institute for Theoretical Physics for hospitality and
the INFN for partial support at a preliminary stage of this work.
BB would like to thank CETUP* (Center for Theoretical Underground Physics and Related Areas), 
supported by the US Department of Energy under Grant No. DE-SC0010137 and by the US National Science 
Foundation under Grant No. PHY-1342611, for its hospitality and partial support during the 2013 Summer Program.
\end{theacknowledgments}



\bibliographystyle{aipproc}   % if natbib is available

\bibliography{sample}

\IfFileExists{\jobname.bbl}{}
 {\typeout{}
  \typeout{******************************************}
  \typeout{** Please run "bibtex \jobname" to optain}
  \typeout{** the bibliography and then re-run LaTeX}
  \typeout{** twice to fix the references!}
  \typeout{******************************************}
  \typeout{}
 }

%%%%%%%%%%%%%%%%%%%%%%%%%%%%%%%%%%%%%%%%%%%
%% The following lines show an example how to produce a bibliography
%% without the help of the BibTeX program. This could be used instead
%% of the above.
%%%%%%%%%%%%%%%%%%%%%%%%%%%%%%%%%%%%%%%%%%%

\end{document}